\documentclass{nrc1}
\usepackage[dvips]{graphicx}
\newcommand\beq{\begin{equation}}
\newcommand\eeq{\end{equation}}
\newcommand\bea{\begin{eqnarray}}
\newcommand\eea{\end{eqnarray}}
\newcommand\beano{\begin{eqnarray*}}
\newcommand\eeano{\end{eqnarray*}}
\newcommand\LL{{\cal L}}
\newcommand\eps{\epsilon}
\newcommand\al{\alpha}
\newcommand\be{{\beta}}
\newcommand\ga{{\gamma}}
\newcommand\om{{\omega}}
\newcommand\ka{{\kappa}}
\newcommand\eqref[1]{(\ref{#1})}

\newcommand{\qb}{$Q$-ball}
\newcommand\nonono[1]{{}}
\journalcode{cjp}
\begin{document}

\title{Maxwell-Chern-Simons Q-balls}

\bigskip\bigskip
\author{M.~Deshaies-Jacques}
\address{Physique des particules, Universit\'e de Montr\'eal,
C.P. 6128, Succ. Centre-ville, Montr\'eal, QC H3C 3J7}
\author{R.~MacKenzie}
\correspond{richard.mackenzie@umontreal.ca}
\shortauthor{M.~Deshaies-Jacques, R.~MacKenzie}
\maketitle

\begin{abstract}
We examine the energetics of $Q$-balls in Maxwell-Chern-Simons theory in
two space dimensions. Whereas gauged $Q$-balls are unallowed in this dimension in the absence of a Chern-Simons term due to a divergent electromagnetic energy, the addition of a Chern-Simons term introduces a gauge field mass and renders finite the otherwise-divergent electromagnetic energy of the $Q$-ball. Similar to the case of gauged $Q$-balls, Maxwell-Chern-Simons $Q$-balls have a maximal charge. The properties of these solitons are studied as a function of the parameters of the model considered, using a numerical technique known as relaxation. The results are compared to expectations based on qualitative arguments.
\end{abstract}

\begin{resume}
Nous examinons l'\'energ\'etique de solitons non topologiques nomm\'es $Q$-balls en th\'eorie Maxwell-Chern-Simons en deux dimensions d'espace. Tandis que les $Q$-balls jaug\'es ne peuvent exister en cette dimension en raison d'une divergence de l'\'energie \'electromagn\'etique, l'ajout du terme de Chern-Simons rend massif le champ de jauge et \'elimine cette divergence. Comme dans le cas jaug\'e, les $Q$-balls Maxwell-Chern-Simons ont une charge maximale. Les propri\'et\'es de ces solitons sont \'etudi\'ees en fonction des param\`etres du mod\`ele, utilisant une technique num\'erique nomm\'ee relaxation. Les r\'esultats ainsi obtenus sont compar\'es aux attentes bas\'ees sur des arguments qualitatifs.
\end{resume}

\section{Introduction\label{sec:intro}}
A class of non-topological solitons (see \cite{Lee:1991ax} for a comprehensive review) dubbed {\qb}s were examined some time ago by Coleman \cite{Coleman:1985ki}. These objects owe their existence to a conserved global charge. Under certain circumstances, a localized configuration of charge $Q$ can be created which has a lower energy than the ``naive" lowest-energy configuration, namely, $Q$ widely-separated ordinary particles (each of unit charge) at zero momentum. This latter state obviously has energy $Qm$, where $m$ is the mass of the quanta of the theory.
If another configuration of charge $Q$ can be constructed whose energy is lower, then that state cannot decay into ordinary matter: either it is stable or some other ``non-naive" configuration of the same charge and still lower energy is stable.

A number of variations have been studied since, including non-abelian {\qb}s \cite{Safian:1987pr,Safian:1988cz}, gauged {\qb}s \cite{Lee:1988ag,Anagnostopoulos:2001dh}, {\qb}s in other dimensions \cite{Prikas:2004fx,Gleiser:2005iq}, higher-dimensional $Q$-objects \cite{MacKenzie:2001av,Axenides:2001pi}, spinning {\qb}s \cite{Volkov:2002aj,Kleihaus:2005me}, and so on.

Of particular interest here is the paper of Lee, et al. \cite{Lee:1988ag}, who discussed the case of gauged {\qb}s in three dimensions, using a combination of analytical and numerical techniques. 

In two dimensions, the gauged \qb's existence is problematic, for a fairly straightforward reason: the electric field goes like $1/r$ and the electric field energy diverges logarithmically. Such a divergence is sufficiently mild that one could still contemplate a configuration of several positively- and negatively-charged {\qb}s with total charge neutrality, in the spirit of global cosmic strings and vortices in liquid helium which also have logarithmically divergent energies. Nonetheless, strictly speaking, an isolated gauged {\qb} in two dimensions has divergent energy and therefore cannot hope to compete energetically with ordinary matter.

However, a new possibility exists in two dimensions: one can consider a model with a Chern-Simons term, either on its own or in addition to the usual Maxwell term. Among the well-known physical effects of the Chern-Simons term (to say nothing of its profound mathematical properties) are a greater interplay between electric and magnetic phenomena (for example, a static charge distribution gives rise to both electric and magnetic fields) \cite{MacKenzie:1988ft,Goldhaber:1988iw}, parity and time reversal violation \cite{Redlich:1983kn}, fractional spin and statistics \cite{Wilczek:1983cy,Arovas:1985yb}, and mass generation for the gauge field \cite{Deser:1981wh}. The latter property is particularly pertinent here since the electric field of a Maxwell-Chern-Simons (MCS) {\qb} decays exponentially, and the argument given above leading to the conclusion that the electric field energy diverges no longer applies. Thus, we can address the question of whether gauged {\qb}s exist (that is to say, whether they can compete energetically with ordinary matter) in 2 space dimensions if the Chern-Simons term is present.

In this paper we do such an analysis, numerically. We begin by briefly reviewing refs. \cite{Coleman:1985ki,Lee:1988ag}, which discuss the ungauged and gauged cases, respectively. We then move on to the MCS case, describing the model studied and the \qb\ ansatz generalized to the MCS case. Next, we make some qualitative observations to indicate what we might expect. In the remainder of the paper, we describe the numerical approach used and the results obtained.

Before getting our hands dirty, it is perhaps useful to address the question: Why should we study MCS {\qb}s in the first place? The motivation lies mainly in the intrinsic interest of the model studied, and of the Chern-Simons term in particular. This term is as interesting mathematically as physically. On the mathematical side, one can mention topological field theory and applications to knot theory; on the physical side, at least two concrete physical applications have been advanced: the fractional quantum Hall effect (see \cite{Prange:1990} for a thorough discussion), and a proposed mechanism of superconductivity based on anyons \cite{Wilczek:1990ik}. Thus, it is worth studying any effect of this term. Furthermore, the model studied below (see \eqref{one}) is a relatively simple and uncontrived one, and whenever a fairly simple model gives rise to interesting phenomena such as solitons, it is worth examining in detail, without the need to evoke concrete physical applications to justify the work.

\section{Ungauged Q-balls\label{sec:ungauged}}
The simplest model in which {\qb}s exist is a complex scalar field theory. A wide variety of potentials give rise to {\qb}s; the precise requirements of the potential are that it be minimized at the origin (so that the symmetry is unbroken), and that a parabola passing through the origin exists which, firstly, is wider than the potential at the origin, and, secondly, intersects the potential at some nonzero field value.

The {\qb} ansatz is $\phi(x)=f(r)\exp(-i\om t)$, where $f(r)$ is a real function interpolating between some nonzero value (determined dynamically) at the origin and zero at infinity. Since the equation of motion for $f$ is a single second-order differential equation, a simple but revealing mechanical analogy exists, wherein $f$ is the position and $r$ the time; with this, one can see that {\qb}s have the following properties:
1. They exist with any charge;
2. As $\om$ increases, the charge of the {\qb} decreases;
3. For large $Q$, $f$ is approximately constant and independent of $Q$ inside the {\qb}, and the energy is proportional to $Q$.
Furthermore, there is no significant difference in 2+1 dimensions (that is to say, all these properties remain true, though the details of the solution will, of course, differ).

\section{Gauged Q-balls\label{sec:gauged}}
The gauged case is considerably more difficult to analyze, essentially because the mechanical analogy mentioned in the previous section is no longer useful. Nonetheless, Lee, et al. \cite{Lee:1988ag} studied the case of gauged {\qb}s in three dimensions. They argued that when the charge exceeds a critical value the {\qb}'s energy exceeds $Qm$, so the \qb\ is at best metastable. This is intuitively reasonable, since a ball of electric charge will have a Coulomb energy which grows roughly as the square of the charge, so eventually the \qb\ will be unable to compete with ordinary matter. On the other hand, as the charge decreases the {\qb} gets smaller and smaller; surface effects become important and eventually destabilize the {\qb}. These two observations indicate that there may or may not be a range of charges for which {\qb}s exist, depending on the point at which each of these effects becomes significant.

The gauged {\qb} ansatz is that of the ungauged case to which is added a scalar potential which describes the electric charge of the {\qb}. In a way, the gauged {\qb} is the compliment of a superconducting vortex: whereas the latter has $|\phi|$ interpolating between zero at the origin and the non-zero vev at infinity (so that the core is normal as opposed to superconducting), the gauged {\qb} is superconducting in the core and normal outside.

As indicated above, while there is no significant difference between {\qb}s in ungauged models in 2+1 and 3+1 dimensions, this is not the case with gauged {\qb}s: they have infinite energy in 2+1 dimensions. This leads us to consider the addition of a Chern-Simons term, with which the energy is rendered finite.

\section{Maxwell-Chern-Simons Q-balls: Generalities\label{sec:mcs:g}}
The model we consider has a complex scalar field $\phi(x)$ with gauged U(1) symmetry, described by the following Lagrangian (in 2+1 dimensions):
\beq
\LL=-\frac14 F_{\mu\nu}^2+{\kappa\over2}\eps^{\al\be\ga}A_\al\partial_\be A_\ga
+|D_\mu\phi|^2
-V(\phi),
\label{one}
\eeq
where $F_{\mu\nu}=\partial_\mu A_\nu - \partial_\nu A_\mu$ and $D_\mu\phi=(\partial_\mu+ie A_\mu)\phi$. The potential must satisfy the same requirements mentioned in Section~\ref{sec:ungauged}. These are satisfied with the potential
\beq
V(\phi)=\phi^*\phi-\frac12(\phi^*\phi)^2+\frac{g}3(\phi^*\phi)^3,
\label{two}
\eeq
if $g>3/16$.

Since the Chern-Simons term renders the gauge field massive, the electric field decays exponentially rather than like $1/r$, and its contribution to the {\qb} energy is finite.
By rescaling the fields and position, the mass of the $\phi$ field can be set to unity, setting the standard to which {\qb}s must be compared. If we can construct a field configuration for which $E/Q<1$ (where $Q$ is the particle number, henceforth referred to simply as charge), then it cannot decay into ordinary matter, and either it or some other lower-energy configuration of the same charge is stable. To look for such a configuration, we use an ansatz much like that of the gauged {\qb}, along with an appropriate form for the vector potential (recall that, with the Chern-Simons term, electric charge is a source for both electric and magnetic fields), namely:
\beq
\phi(x)=e^{-i\om t}f(r),\quad
A^0(x)=\al(r),\quad
A^i(x)= {\eps^{ij}r_j\over r}\beta(r).
\label{five}
\eeq
%%%%%%%%%%
\nonono{
The field equations become
\bea
f'' + {f'\over r}+\left((\om-e\al)^2-e^2\beta^2 -1\right)f +f^3-gf^5&=&0,\nonumber\\
\al''+{\al'\over r}-\kappa(\be'+{\be\over r})+2e(\om-e\al)f^2&=&0,
\label{six}\\
\be''+{\be'\over r}-{\be\over r^2}-\kappa \al' -2e^2\be f^2&=&0.\nonumber
\eea

\noindent The boundary conditions at the origin are
\beq
f'(0)=0,\quad \al'(0)=0,\quad \be(0)=0,
\label{eight}
\eeq
while as $r\to\infty$ these three fields must tend towards zero in the following way:
\beq
f(r)\sim e^{-\sqrt{1-\om^2}r},\quad \al(r)\sim e^{-\ka r},\quad
\be(r)\sim 1/r.
\label{nine}
\eeq
The asymptotic form of $\be$ is a pure gauge, and describes the total magnetic flux, which can be seen to be proportional to the charge by integrating the second of Eqs. \eqref{six}.}

The fields $f$ and $\al$ are nonzero at the origin and tend to zero exponentially as $r\to\infty$, while $\be$ is zero at the origin, rises, and then returns to zero (as $1/r$) as $r\to\infty$.

There is a restriction on the frequency $\omega$ for which {\qb}s may exist. The $\omega$-dependent term can be thought of as being a part of an effective potential, and (as explained in \cite{Coleman:1985ki}) for {\qb}s to exist,
\beq
\sqrt{1-{3\over16g}}<\om<1.
\label{ten}
\eeq
For ungauged {\qb}s, values of $\om$ near the upper end of this range correspond to small {\qb}s, while the lower end of the range corresponds to large {\qb}s. Things are slightly more complicated in the gauged case (whether with or without a Chern-Simons term), as will be described below.

The equations of motion have four parameters: $e,\ g,\ \ka,\ \om$. Of these, the first three are parameters of the model itself (appearing in \eqref{one}), while the fourth is a parameter of the ansatz. 
%%%%%%%%%%
\nonono{A potentially interesting limit of the model is the pure Chern-Simons case. This can be realized by the change of variables $\bar A_\mu\equiv e A_\mu$ and $\bar\kappa\equiv\kappa/e^2$, after which the only appearance of $e$ in the Lagrangian is in the first term, which becomes $-\bar F_{\mu\nu}^2/4e^2$. The Maxwell term is then eliminated by setting $e\to\infty$ with $\bar\kappa$ fixed. We have not studied this limit here, although our analysis suggests that {\qb}s would require a very large value of $\bar\kappa$.}

Once a solution is found, the energy and charge are evaluated numerically in a straightforward manner, and the ratio of the two indicates whether decay to ordinary matter is energetically possible or not.
Of particular interest is the lowest value of $E/Q$ for given parameters of the model, since this indicates the maximal energy savings gained in forming a {\qb}; it is thus an indicator of the most stable configuration (at least, among {\qb}s).

%%%%%%%%%%
\nonono{Given that the model has three parameters, a complete exploration of parameter space would be rather involved. Rather than do this, since we are most interested in the effect of the Chern-Simons term on the existence and properties of {\qb}s, we have restricted ourselves to a specific value of $g$ (namely, $g=0.5$), chosen arbitrarily, apart from the fact that it does fall into the allowed range given above.}

What can be said qualitatively, before attacking the problem numerically? Even in a gauged model without Chern-Simons term \cite{Lee:1988ag}, a qualitative analysis is much more difficult than in the ungauged case because, with the addition of the gauge field, Coleman's mechanical analogy no longer applies. This is all the more so in the MCS case, where the ansatz has three fields. However, a couple of general observations can easily be made. 

Suppose we held $Q$ fixed and ``turned on" $e$. Clearly, this would create an additional contribution to the {\qb} energy coming from the electromagnetic field. Thus, we would expect $E/Q$ to increase with $e$ for fixed $Q$; furthermore, we would expect the maximal charge at which {\qb}s occur to decrease with increasing $e$, very similar to the gauged case \cite{Lee:1988ag}.

As for the Chern-Simons term, as mentioned above, its presence is essential (in two dimensions) since without it the {\qb}'s energy diverges, so we cannot turn its coefficient $\ka$ on, since it cannot be zero. How do we expect the energetics to vary as the coefficient of the Chern-Simons term varies? Since the gauge field's mass 
is proportional to $\ka$, the electromagnetic fields have a decay length of $1/\ka$, so as $\ka$ increases, the electromagnetic contribution to the {\qb} energy should decrease, and {\qb}s should be stable over a wider range of the other parameters (or, equivalently, over a wider range of charges). 

These two qualitative behaviours are indeed seen in our numerical work, as will now be discussed.

\section{Maxwell-Chern-Simons Q-balls: Numerical analysis\label{sec:mcs:n}}
The search for {\qb}s was performed using an iterative method known as relaxation, as described in detail in \cite{Press:1992}. A discretized configuration is provided as an initial guess; the algorithm estimates an error by determining to what extent this configuration is {\em not} a solution, adds a correction to the configuration in an intelligently-chosen ``direction" in configuration space, re-estimates the error, and so on, until the error is sufficiently small.

%%%%%%%%%%
\nonono{Note that the relaxation process is in no sense an actual physical evolution of the system. In particular, the charge is not conserved from one iteration to the next. The method attempts to find an approximate solution, given the four parameters $e,\ g,\ \ka,\ \om$. A more physical approach would be to specify $e,\ g,\ \ka$ and the charge $Q$ and calculate the minimum-energy solution for fixed $Q$; however, $Q$ and $\om$ are inextricably linked, and we could not replace one by the other.}

Let us describe in detail our results for typical values of the parameters of the model: $(e,g,\ka)=(0.1,0.5,2.0)$. Varying $\om$ within the allowed range given in \eqref{ten} (for $g=0.5$, this is $0.7906\le\om\le1$) reveals two classes of {\qb}s, which can be described as small and large {\qb}s. Small {\qb}s, with charges ranging from about 20 to roughly 2000, exist in the range $0.8111\leq\om<1.0$, while large {\qb}s, with charges ranging from about 2000 to 43000, exist for $0.8111\leq\om\leq0.8995$. For $\om$ in the region $0.7906\le\om\le0.8111$, no {\qb} solutions were found.

\begin{figure}[ht]
\begin{center}
\begin{tabular}{cc}
\includegraphics[width=7cm]{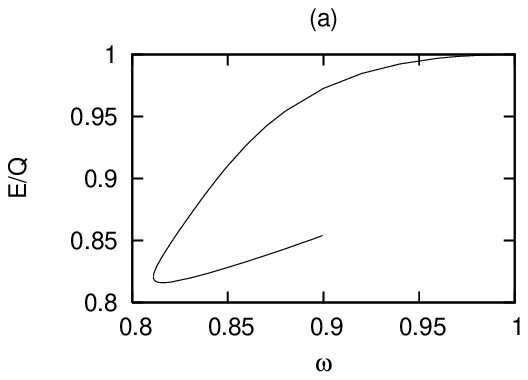} &
\includegraphics[width=7cm]{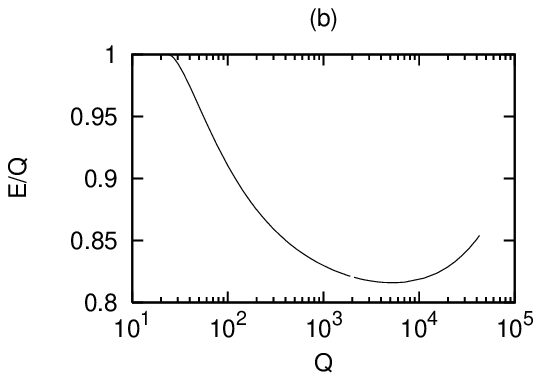} \\
\includegraphics[width=7cm]{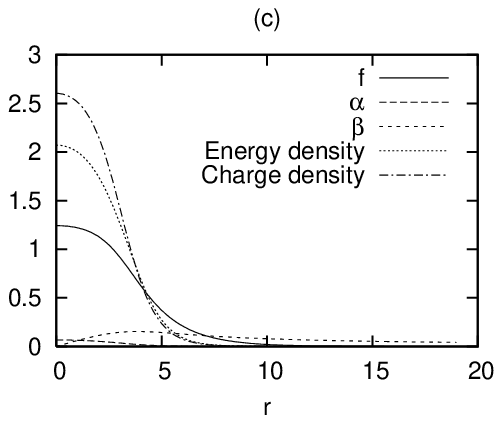} &
\includegraphics[width=7cm]{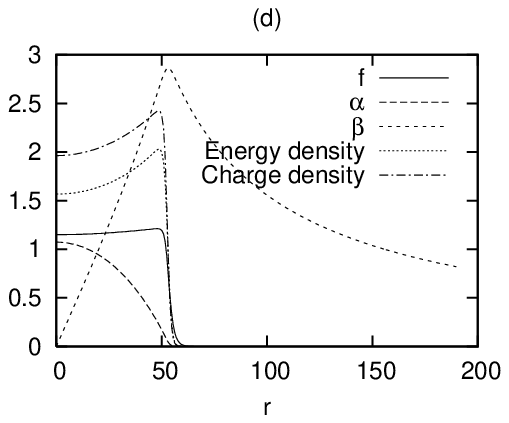}
\end{tabular}
\end{center}
\caption{$E/Q$ vs (a) $\om$ and (b) $Q$, at parameter values $(e,g,\ka)=(0.1,0.5,2.0)$. Note the tiny gap between the upper and lower branches of (a) (corresponding to small  and large {\qb}s, respectively), and the left and right branches of (b) (idem). Typical (c) small and (d) large {\qb} profiles, at $\om=0.85$. Their charges are 99.998 and 19557, respectively.}
\label{figone}
\end{figure}

Since $E/Q\ge1$ for ordinary matter, the energetic advantage of forming a {\qb} can be seen by comparing $E$ and $Q$. Fig.~\ref{figone} shows $E/Q$  as a function of $\om$ and of $Q$. The tiny gap in these figures is the numerical artifact just mentioned, due to the numerically delicate transition between small and large {\qb}s.

Also displayed in Fig.~\ref{figone} are typical small and large {\qb}s. Note that, as was the case with gauged {\qb}s \cite{Lee:1988ag}, the charge density of large {\qb}s increases from the centre to the exterior before dropping off, a behaviour which can be attributed to the repulsive gauge force. Note also that the prominent ``tail" of the field $\beta$ in Fig.~\ref{figone}d is a pure gauge, as mentioned above; all physical quantities tend to zero exponentially as $r\to\infty$, as expected.

%%%%%%%%%%
\nonono{The distinction between {\em small} and {\em large} {\qb}s is not arbitrary. Indeed, note the substantial overlap of the regions where small and large {\qb}s are found: for $0.8111\leq\om\leq0.8995$, this was the case.
For any $\om$ in this overlap region, {\qb}s of two different charges exist. For instance, at $\om=0.85$, the charges turn out to be very close to 100 and 20000. As $\om$ decreases, the difference between the charges decreases, tending toward zero as $\om$ approaches 0.8111, though numerical instability left a tiny gap (which we believe to be a numerical artifact) between the two.

\begin{figure}[ht]%
\begin{center}
\begin{tabular}{cc}
\includegraphics[width=7cm]{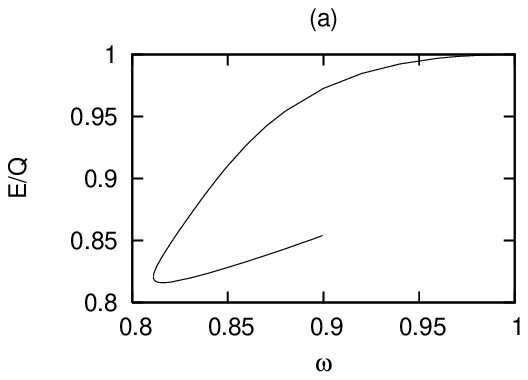} &
\includegraphics[width=7cm]{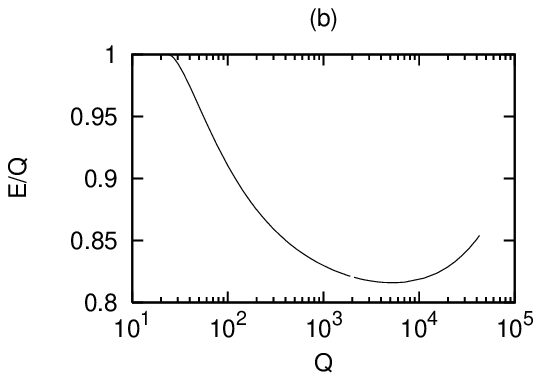}
\end{tabular}
\end{center}
\caption{$E/Q$ vs (a) $\om$ and (b) $Q$, at parameter values $(e,g,\ka)=(0.1,0.5,2.0)$. Note the tiny gap between the upper and lower branches of (a) (corresponding to small  and large {\qb}s, respectively), and the left and right branches of (b) (idem).}
\label{figtwo}
\end{figure}
}

%%%%%%%%%%
\nonono{Nothing particularly dramatic occurs in the extreme cases; for instance, as $\om$ approaches unity (the maximum allowed value), the {\qb} profile tends smoothly towards a nontrivial but qualitatively unremarkable limit; its charge approaches the smallest {\qb} charge found, namely, about 23.63. In this same limit, the energetic advantage, $E/Q$, approaches 1 from below, so these {\qb}s are only marginally preferable over ordinary matter. Similarly, at the large end of the large {\qb} curve, no dramatic change occurred to the solution found; from one value of $\om$ to the next the solution simply disappeared. No amount of coaxing (for instance, changing $\om$ extremely slowly and using the previous solution as initial guess) could entice the program to converge. We also attempted to study these endpoints using a different method (shooting, also described in \cite{Press:1992}), to no avail.}

Two features of our results are unexpected. The first concerns the fact that, while $Q$ is a monotonic decreasing function of $\om$ in the ungauged case, this is not the case of MCS {\qb}s: small {\qb}s act in this way, but the charge of large {\qb}s is a monotonic {\em increasing} function of $\om$. Indeed, the very existence of two {\qb}s at the same value of $\om$ is unlike the ungauged case.

The second unexpected feature of our results concerns the maximum {\qb} charge. As explained above, while there is no upper limit to the ungauged {\qb}'s charge, we anticipate a maximal charge in the gauged case (with or without the Chern-Simons term), due to the electromagnetic contribution to the {\qb}'s energy. This contribution, one would think, should give rise to $E/Q>1$, at which point {\qb}s are no longer energetically advantageous compared with ordinary matter. Indeed there {\em is} a maximal charge. However, as can be seen from Fig.~\ref{figone}a, this occurs when $E/Q$ is increasing but nonetheless considerably less than one. (At the maximal charge, $E/Q=0.8540$). This is also the case with gauged {\qb}s in three dimensions, as can be seen from Fig.~3 of \cite{Lee:1988ag}. Perhaps there is another explanation for this maximal charge (other than the inability to compete energetically with ordinary matter), but we have not come up with one. 

Varying $e$ and $\ka$ had the anticipated effect on {\qb} energetics, as described in the previous section. Details can be found in Ref. \cite{Deshaies-Jacques:2006ae}.
%%%%%%%%%%
\nonono{We are now in a position to speculate on what is to be expected in the pure Chern-Simons limit. This corresponds to $e\to\infty$ with $\kappa/e^2$ fixed. Figures \ref{figthree} and \ref{figfour} indicate that {\qb}s offer the greatest energy advantage for small $e$ and large $\kappa$; these suggest that $\bar\kappa$ must be quite large in order for {\qb}s to be energetically advantageous in this limit. This speculation could be tested with further work.}

\section{Conclusions\label{sec:concl}}
To summarize, we have argued that in two space dimensions, a Chern-Simons term (or some other mass generation mechanism for the gauge field) is necessary in order for {\qb}s to have finite energy. Considering the simplest model which might then give rise to {\qb}s, we have performed a numerical search for these objects. Two types were found for a wide range of parameters, large and small {\qb}s, the former being more stable in that they typically have lower values of $E/Q$.  While many aspects of the behaviour of these objects are largely as expected, the fact that large {\qb}s cease to exist when they do (in particular, with $E/Q$ well below unity) is surprising and merits further study.

\bigskip
We thank Manu Paranjape for interesting conversations. This work was
funded in part by the
National Science and Engineering Research Council.

\bibliographystyle{unsrt}
\bibliography{fieldtheory}

\end{document}